\documentclass[amsmath,amssymb,aps,letterpaper,prl,twocolumn,longbibliography,superscriptaddress]{revtex4-2}
\usepackage{tikz}
\usepackage[english]{babel}
\usepackage[utf8]{inputenc}
\usepackage[T1]{fontenc}
\usepackage{indentfirst}
\usepackage{amsmath}
\usepackage{amssymb}
\usepackage{amsthm}
\usepackage{proof}
\usepackage{graphicx}

\allowhyphens

\newcommand{\mA}{\mathcal{A}}
\newcommand{\mS}{\mathcal{S}}
\newcommand{\mD}{\mathcal{D}}
\newcommand{\mB}{\mathcal{B}}

\newcommand{\valos}{\mathbb{R}}

\newcommand{\ket}[1]{{\left|#1\right\rangle}}
\newcommand{\bra}[1]{{\left\langle #1\right|}}

\newcommand{\x}{X}
\newcommand{\y}{Y}
\newcommand{\z}{Z}


\usepackage{ifpdf}

\ifpdf
\usepackage{epstopdf}
\usepackage[pdftex,colorlinks,urlcolor=blue,citecolor=blue,linkcolor=blue]{hyperref}
\else
\usepackage[hypertex,colorlinks,urlcolor=blue,citecolor=blue,linkcolor=blue]{hyperref}
\fi
\begin{document}

\title{
Sub-lattice entanglement in an exactly solvable anyon-like spin ladder
}

\author{Bal\'azs Pozsgay}
\affiliation{MTA-ELTE “Momentum” Integrable Quantum Dynamics Research Group, Department of Theoretical Physics, Eötvös
  Loránd University, P\'azm\'any P\'eter stny. 1A, Budapest 1117, Hungary}
\author{Arthur Hutsalyuk}
\affiliation{MTA-ELTE “Momentum” Integrable Quantum Dynamics Research Group, Department of Theoretical Physics, Eötvös
  Loránd University, P\'azm\'any P\'eter stny. 1A, Budapest 1117, Hungary}
\author{Levente Pristy\'ak}
\affiliation{MTA-ELTE “Momentum” Integrable Quantum Dynamics Research Group, Department of Theoretical Physics, Eötvös
  Loránd University, P\'azm\'any P\'eter stny. 1A, Budapest 1117, Hungary}
\affiliation{Department of Theoretical Physics, Institute of Physics, Budapest University of Technology and Economics, Műegyetem rkp. 3., H-1111 Budapest, Hungary}
\author{G\'abor Tak\'acs}
\affiliation{Department of Theoretical Physics, Institute of Physics, Budapest University of Technology and Economics, Műegyetem rkp. 3., H-1111 Budapest, Hungary}
\affiliation{MTA-BME Quantum Dynamics and Correlations Research Group, Budapest University of Technology and Economics, Műegyetem rkp. 3., H-1111 Budapest, Hungary}

\begin{abstract}
We introduce an integrable spin ladder model and study its exact solution, correlation functions, and entanglement
properties. The model supports two particle types (corresponding to the even and odd sub-lattices), such that the
scattering phases are constants: particles of the same type scatter as free fermions, whereas the inter-particle phase
shift is a constant tuned by an interaction parameter. Therefore, the spin ladder bears similarities with
anyonic models. We present exact results for the spectrum and correlation functions, and we study the sub-lattice
entanglement  by  numerical means. 
\end{abstract}

\maketitle

\section{Introduction}

Classical simulations of quantum many body systems are limited by the growth of entanglement, both in equilibrium and
out-of-equilibrium situations. This motivates the study of models where exact solutions can be found, at least for
certain physical quantities. Important classes of solvable many body systems are the free theories, the one dimensional
integrable models \cite{sutherland-book,Korepin-book}, and also the recently discovered dual unitary quantum circuits
\cite{dual-unitary-4}. 

Integrable models have been studied extensively over many decades, and in the last 10 years their non-equilibrium
dynamics also received considerable attention. Now it is understood that isolated integrable models equilibrate to the
Generalized Gibbs Ensemble \cite{rigol-quench-review,essler-fagotti-quench-review}, and their transport properties are
described by Generalized Hydrodynamics \cite{ghd-review-intro}. However, these results describe only the large time limit for large system sizes, and generally they lack a completely rigorous
proof. Furthermore, they don't provide access to certain exotic features of the dynamics, such as anomalous current
fluctuations \cite{znidaric-anomalous,anomalous-contin,prosen-MM-anomalous,box-ball-currents,vasseur-anom}.

This motivates the study of selected integrable models with even simpler dynamics, where there is some
interaction in the system, nevertheless closed form results
can be derived for the real time evolution of certain physical quantities. Such models include the Rule54
cellular automaton 
\cite{rule54,rule54-review,rule54-op-entangl,rule54-entangl,katja-bruno-rule54-ghd,katja-bruno-lorenzo-rule54}, 
the box-ball systems \cite{box-ball,box-ball-ghd,box-ball-currents}, 
classical cellular automata of the XXC type
\cite{XXC,prosen-MM1,prosen-MM2,prosen-MM-anomalous,sajat-superintegrable,marko-XXC},
non-trivial strong coupling limits of known models \cite{sajat-qboson,sajat-q2,bruno-hubbard}
including the folded XXZ model
\cite{folded1,folded2,sajat-folded,folded4}, or quantum circuits that are both integrable and dual-unitary
\cite{lorenzo-abanin}. A common property of these models is that the scattering of the particles (either classical or
quantum) is rather simple compared to a generic integrable model.

What are the simplest possible interacting $S$-matrices for integrable (Hermitian) quantum spin
chains? An especially 
simple case is when the $S$-matrix is constant (momentum-independent). For a single particle species the only
possibilities are phase factors 
$\pm 1$ corresponding to free bosons/fermions. However, for multiple particle species we can explore a wider range of
options.

In this paper we introduce a new spin ladder model, which supports two particle types which propagate on
the two legs of the spin ladder.  
In this model the scattering phases
are constant statistical factors. This bears strong similarity with anyonic models
\cite{rigol-anyon,anyon-ground,cooper-anyon,kundu-anyon,patu-korepin-anyon}, or parafermionic chains
\cite{free-parafermion,anyonic-tight,gritsev-parafermion}. A crucial difference is that in our model the anyon-like
phases arise from a local interaction defined in the standard spin basis. This provides a unique opportunity to study
the entangling effects of constant scattering phases.

\section{The model and its integrability}

We consider a spin chain made of qubits, using the notation $X_j$, $Y_j$ and $Z_j$ for operators given by the Pauli
matrices, acting on site $j=1\dots L$, where $L$ (even) is the length of the chain.

\begin{figure}[h]
  \centering
  \includegraphics[scale=0.6]{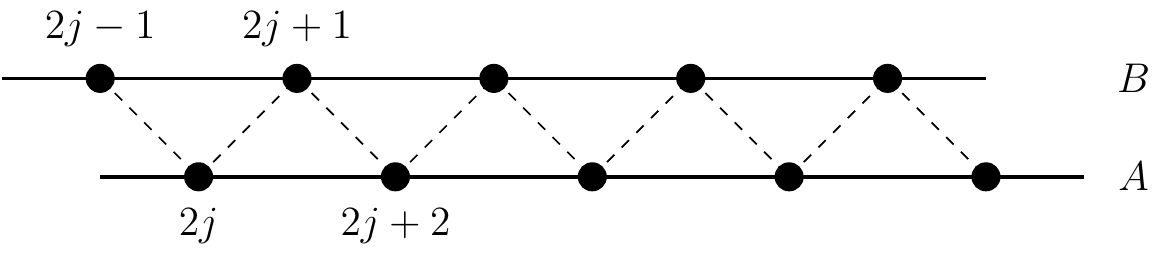}
  \caption{The geometry of the spin zigzag ladder. $A$ and $B$ denote the two sub-lattices on which the particles can hop.}
  \label{fig:1}
\end{figure}

Our model can be seen as a spin chain or as a spin ladder in a zig-zag geometry, see Fig. \ref{fig:1}. We consider a
hopping model on the 
ladder, such that the two legs have a minimal coupling between them. Particles can propagate on the two legs separately,
but the local hopping phases on one leg also depend on the local occupation numbers on the other leg, leading to a
model Hamiltonian with three site interaction: 
\begin{equation}
  H(\gamma)=\sum_{j=1}^{L/2} h_{2j,2j+1,2j+2}(\gamma)+h_{2j+1,2j+2,2j+3}(-\gamma).
\end{equation}
Here $h_{a,b,c}(\gamma)$ is the Hamiltonian density with a real coupling constant $\gamma\in\valos$,
given by
\begin{equation}
  \label{hdef2}
 h_{1,2,3}(\gamma)=-\big[\sigma^-_1 e^{i\gamma \z_2} \sigma^+_3+\sigma^+_1 e^{-i\gamma \z_2} \sigma^-_3\big],
\end{equation}
where $\sigma^\pm_j=(\x_j\pm i\y_j)/2$ are the standard raising/lowering operators. The Hamiltonian is invariant with
respect to a global spin-flip in the $Z$-basis, and it is also space reflection invariant for reflections that also
exchange the two sub-lattices.

In the following interpret the up spins as a vacuum and the down spins as particles.
The Hamiltonian generates hopping on the two sub-lattices, so that the
sub-lattice magnetizations 
\begin{equation}
    S^A=\sum_{j=1}^{L/2} \z_{2j},\qquad S^B=\sum_{j=1}^{L/2} \z_{2j+1}
\end{equation}
are separately conserved. Here and in the following $A$ and $B$  stand for the even and odd sub-lattices,
respectively. Furthermore the hopping phase is $e^{\pm
  i\gamma}$ depending on the position and the occupation of the sites involved.

To our best knowledge this model has not yet appeared in the literature. Similar models include the so-called extended
XX model \cite{suzuki-xy,extended-xy}, a super-symmetric hopping model treated in \cite{cooper-anyon}, or the Bariev model
\cite{bariev-model}. However, our model is different, as the exact solution below shows.

The model has two free fermion points, with different physical interpretation. For $\gamma=0$ the interaction between the two
legs disappears, leading to two uncoupled XX spin chains. For $\gamma=\pi/2$  the model becomes a special case of the
extended XX models \cite{extended-xy}, which can be solved by a single Jordan-Wigner 
transformation, see below.

The model is integrable for any coupling $\gamma$. It has an infinite set of conserved charges, which are given by a
diagonal dressing of the known conserved charges of the XX spin chains \cite{GM-higher-conserved-XXZ}. 
The charges are organized into four infinite families, corresponding to the two sub-lattices and two ``chiralities''.
The charges can be expressed in terms of densities as
\begin{equation}
  Q^{A+}_\alpha=\sum_{j=1}^{L/2} q^{A+}_\alpha(2j),\qquad Q^{B+}_\alpha=\sum_{j=1}^{L/2} q^{B+}_\alpha(2j+1),
\end{equation}
together with $Q^{A-}_\alpha=(Q^{A+}_\alpha)^\dagger$ and $Q^{B-}_\alpha=(Q^{B+}_\alpha)^\dagger$. The index $\alpha$
denotes the range of the given operator density.

The shortest charge densities appear for $\alpha=3$ and they are simply just terms from the Hamiltonian:
\begin{equation}
  \begin{split}
    q^{A+}_3(2k)&= \sigma^+_{2k}D^\dagger_{2k+1}  \sigma^-_{2k+2}\\
 q^{B+}_3(2k+1)&=  \sigma^+_{2k+1}D_{2k+2}   \sigma^-_{2k+3}, \\
  \end{split}
\end{equation}
where we defined 
\begin{equation}
D_j=e^{i\gamma \z_j}=\cos(\gamma)+i\sin(\gamma)\z_j. 
\end{equation}
Higher charges are constructed by a mixed diagonal dressing of the hopping terms of the type
$\sigma^+_k\sigma^-_{k+\alpha-1}$. The dressing is such that for each site between $k$ and $k+\alpha-1$ we include a
$\z$ operator if the site is from the same sub-lattice, and a $D$ operator otherwise. For example, for range
$\alpha=5$ we get
\begin{equation}
  \begin{split}
    q^{A+}_5(2k)&= \sigma^+_{2k}D^\dagger_{2k+1} \z_{2k+2} D^\dagger_{2k+3}   \sigma^-_{2k+4}\\
 q^{B+}_5(2k+1)&=  \sigma^+_{2k+1}D_{2k+2} \z_{2k+3} D_{2k+4}   \sigma^-_{2k+5}, \\
  \end{split}
\end{equation}
together with $q^{A-}_5(2k)=\left(q^{A+}_5(2k)\right)^\dagger$ and $q^{B-}_5(2k+1)=\left(q^{B+}_5(2k+1)\right)^\dagger$. Higher
charges can be constructed in 
an analogous way. The commutativity of all of these charges is proven as follows. First of all, all charges belonging to either sub-lattice $A$ or $B$ necessarily commute with each other, because they are just diagonal dressings of the
charges of an XX model localized on one of the sub-lattices \cite{GM-higher-conserved-XXZ}. The commutativity of the
charges corresponding to different 
sub-lattices is less obvious. However, in this case the charge densities actually commute, for example
direct computation gives
\begin{equation}
[ q^{A+}_3(2k),q^{B+}_3(2l+1)]=0
\end{equation}
for all $k,l$. A more complete proof can be given using a similarity transformation discussed below.

The existence of an infinite family of commuting charges implies that the model is integrable and it has 
a completely elastic and factorized $S$-matrix \cite{kulish-s-matrix,Mussardo-review}. We derive this $S$-matrix
below. The charges above can be embedded into a transfer matrix constructed from local Lax operators, using the
framework of \cite{sajat-medium}, but we do not treat this approach here.

\section{The solution of the model}

For $\gamma=0$ the model can be solved by two independent Jordan-Wigner (JW) transformations performed on the two
sub-lattices. It is then a natural idea to construct a generalized JW transformation also for finite $\gamma$. To this
order let us consider the model with free boundary conditions (or alternatively, a half-infinite chain). We introduce
creation and annihilation operators for the two sub-lattices as
\begin{equation}
  \begin{split}
    c^{A}(2j)&=     D_1     \z_2   D_3   \z_4  \dots \z_{2j-2} D_{2j-1} \sigma^+_{2j}\\
   c^{B}(2j+1)&=\z_1  D_2^\dagger  \z_3  D_4^\dagger  \dots     \z_{2j-1}  D_{2j}^\dagger   \sigma^+_{2j+1},    \\
  \end{split}
\end{equation}
together with their adjoints.

Direct computation shows that
\begin{equation}
  \label{comm1}
  \{c^A(2j),c^A(2k)\}=0,\quad
  \{c^A(2j),c^{A\dagger}(2k)\}=\delta_{jk} 
\end{equation}
and similarly for the $B$ sub-lattice. However, for the cross-commutation terms we get for example
\begin{equation}
  \label{comm3}
  c^{A}(2j)c^B(2k+1)=e^{2i\gamma}  c^B(2k+1)c^A(2j),\quad j<k.
\end{equation}
This means that the model is partially anyonic. The Hamiltonian can be rewritten for all $\gamma$ 
as
\begin{equation}
  \label{Hc}
  H(\gamma)=-\sum_{j=1}^{L/2-1}\big[ (c_{2j+2}^A)^\dagger c_{2j}^A+(c_{2j+3}^B)^\dagger c_{2j+1}^B\big]+c.c.
\end{equation}
This means that all the interaction is included now in the definition of the creation/annihilation operators. This is
similar to parafermionic models \cite{free-parafermion,anyonic-tight,gritsev-parafermion}, but in our case the
non-trivial commutation relations arise from the local interaction in the model and not from pre-defined operator algebras.

We recover the usual JW transformation in two special cases. If $\gamma=0$ then we get two independent
JW transformations on the two sub-lattices. In contrast, for  $\gamma=\pi/2$ we
get a single JW transformation on the whole chain. Thus the model interpolates between two uncoupled XX chains and a
single XX chain, although in the latter case the Hamiltonian is actually a higher charge of the usual XX model
\cite{extended-xy}. 

While for the free cases the model can be solved by Fourier transform, this does not work for generic $\gamma$ due to
the mixed anyon-like interactions. In these cases we can use the fact that the phase factors do not depend
on the momenta of the particles, only on the relative position of the particles on the two 
sub-lattices. This leads to a simple explicit construction of the wave functions.

Let us consider a state with $N_A$ and $N_B$ particles on the two sub-lattices. Lattice momenta of the particles will be
denoted as $p^{A}_j$ and $p^B_k$. For writing down the wave function we will use coordinates ${\bf a}=\{a_j\}_{j=1,\dots,N_A}$ and
${\bf b}=\{b_k\}_{j=1,\dots,N_B}$ which run
over the even and odd numbers, respectively.
First we consider periodic boundary conditions.
The wave function is then given by 
\begin{equation}
      \label{bethestate}
  \Psi({\bf a},{\bf b})=\det(\mA) \det(\mB) \prod_{a<b} e^{-i\gamma} \prod_{b<a} e^{i\gamma}.
\end{equation}

Here $\mA$ and $\mB$ are matrices of sizes $N_A\times N_A$ and $N_B\times N_B$, respectively. They describe free
fermionic wave functions localized on the two sub-lattices, with components given by 
\begin{equation}
  \mA_{jk}=e^{i(p^A_j-\gamma) a_k/2},\qquad   \mB_{jk}=e^{i(p^B_j+\gamma) b_k/2}.
\end{equation}
The interpretation of this wave function is the following: The model supports two particle species
moving on the two 
sub-lattices, with momenta $p^A_j$ and $p^B_j$. 
The scattering in the model is factorized and diagonal, with
momentum independent phase shifts given by
\begin{equation}
  \mS_{AA}=\mS_{BB}=-1,\qquad \mS_{AB}=\left(\mS_{BA}\right)^{-1}=e^{2i\gamma}.
\end{equation}
The phase shifts reflect the commutation relations \eqref{comm1}-\eqref{comm3}.

Periodicity implies that the momenta have to satisfy the Bethe equations
\begin{equation}
     \label{Betheeq}
  \begin{split}
  e^{ip^A_j L/2} &=(-1)^{N_A-1} e^{i\gamma (L/2-2N_B)}\\
  e^{ip_j^B L/2}&=(-1)^{N_B-1} e^{-i\gamma (L/2-2N_A)}.
  \end{split}
\end{equation}

These equations are almost free: the only coupling between the two sets of momenta is simply just a twist, which depends
on the overall particle numbers. The energy eigenvalues are then
\begin{equation}
  E=\sum_{j=1}^{N_A}e(p^A_j)+\sum_{j=1}^{N_B}e(p^B_j),
\end{equation}
with $e(p)=-2\cos(p)$.

Now we consider the model with free boundary conditions, and show that in this case the wave functions are found simply
using a global similarity transformation. 
The diagonal operator
\begin{equation}
  \mD=\prod_{2j<2k+1} e^{i\gamma Z_{2j}  Z_{2k+1}/4}
  \prod_{2j>2k+1} e^{-i\gamma Z_{2k+1} Z_{2j}/4}
\end{equation}
completely decouples the two legs of the ladder:
\begin{equation}
  \label{simi}
  \mD H(\gamma) \mD^{-1}=H(0).
\end{equation}
This implies that the spectrum of the open chain is the same as that of two uncoupled XX chains
for all $\alpha$. In this sense the model is free, and it belongs to the class of models investigated in \cite{fendley-fermions-in-disguise,fermions-behind-the-disguise}.
Nevertheless, the operator $\mD$ is highly non-local, and it makes the two legs of the ladder highly entangled, both
in equilibrium and out-of-equilibrium situations. 

We also study the thermodynamic limit (TDL).  We introduce root densities $\rho^A(p)$ and $\rho^B(p)$,
and the limiting value of the ground state energy density becomes
\begin{equation}
 \lim_{TDL} \frac{E}{L}=\int_{-\pi}^\pi \frac{dp}{4\pi} e(p)(\rho^A(p)+\rho^B(p)).
\end{equation}
The ground state is given by the half filled state
\begin{equation}
  \rho^{A,B}(p)=
  \begin{cases}
    1 & \text{ for } |p|<\pi/2\\
    0 & \text{ for }  |p|>\pi/2.\\
  \end{cases}
\end{equation}
The ground state energy density is $-2/\pi$, which is identical to that of the XX model.

\section{Correlation functions}

The coupling between the two sub-lattices makes them
strongly entangled. We demonstrate this by computing a selected short range correlation function. Due to
the similarity transformation \eqref{simi} the correlation functions of $Z$ operators will be the same as in two uncoupled XX
chains. Therefore, non-trivial information is seen in correlation functions with hopping terms.
We choose the following connected correlation function:
  \begin{equation}
      \label{Cpsi}
 C_\Psi\equiv  \bra{\Psi}\sigma^-_{0}\z_{1}\sigma^+_{2} \ket{\Psi}-
   \bra{\Psi}\sigma^-_{0}\sigma^+_{2} \ket{\Psi} \bra{\Psi}\z_{1} \ket{\Psi}.
\end{equation}
Here $\ket{\Psi}$ is an arbitrary eigenstate  with $N_A$ and $N_B$ particles on the two sub-lattices. A non-zero value
of $C_\Psi$ demonstrates the entanglement between the two sub-lattices.

We introduce the magnetization on the odd sub-lattice:
\begin{equation}
m_B= \bra{\Psi}\z_{1} \ket{\Psi} =  \frac{L-4N_B}{L}.
\end{equation}
A certain combination of the operators above is simply the density of a conserved charge (one term in the Hamiltonian),
therefore we get
\begin{equation}
   \bra{\Psi}  \cos(\gamma)\sigma^-_{0}\sigma^+_{2}+i\sin(\gamma) \sigma^-_{0}\z_{1}\sigma^+_{2} \ket{\Psi}
    =2W,
\end{equation}
with
\begin{equation}
  W=\frac{1}{L}\sum_{j=1}^{N_A} e^{ip^A_j}.
\end{equation}
Now we apply the Hellmann-Feynman theorem for the corresponding charge, from which we can obtain the mean values of the
$\gamma$-derivative of the operators on the l.h.s. above. Combining this with the mean value above and with the Bethe
equations 
\eqref{Betheeq} we get the result
\begin{equation}
  C_\Psi= 2i  \sin(\gamma)W(m^2_B-1).
\end{equation}

In the thermodynamic limit we get
\begin{equation}
  W\to \int_{-\pi}^\pi \frac{dp}{4\pi} \rho^A(p)e^{ip}.
\end{equation}
This retains a finite value unless the root distribution is constant, therefore we obtain a finite
correlation between the sub-lattices for almost all states.

The correlation function vanishes if either lattice is fully polarized, having magnetization equal to $\pm 1$. If the
odd sub-lattice is polarized, then the vanishing is guaranteed by the factor $(m_B^2-1)$, whereas if the even
sub-lattice is completely polarized, then $W=0$ (because for a fully polarized state $\rho$ is constant).

Analogous results can be obtained for combinations similar to \eqref{Cpsi}.

\section{Sub-lattice entanglement}

We also study the entanglement properties of the model, in both equilibrium and out-of-equilibrium situations.
In the literature the most often studied entanglement is that of connected sub-systems
\cite{cardy-calabrese,pasquale-cardy-entanglement,fagotti-calabrese-entangl,pasquale-vincenzo-PNAS,vincenzo-renyi3,entanglement-spreading-review}. However, it is
expected that in this model the usual bipartite entanglement behaves very
similar to that of the XX model. Therefore we focus on the sub-lattice entanglement, which is a highly non-trivial quantity
that can be tuned by the coupling constant $\gamma$.
Sub-lattice entanglement was studied earlier in a number of situations
\cite{sublattice-entangl-phase1,sublattice-entangl-comb,sublattice-entangl-phase2,alracs-entangl,sublattice-entangl-even-odd,sublattice-entangl-periodic}.

As before, let $A$ and $B$ denote the sites of the even and odd sub-lattices, and we define the sub-lattice
(von Neumann) entanglement entropy as 
\begin{equation}
  \label{Sdef}
  S=-\text{Tr}\big(\rho_A\log\rho_A\big),
\end{equation}
where $\rho_A=\text{Tr}_B \rho$, with $\rho$ being the density matrix of the system, either in equilibrium or in an
out-of-equilibrium process. The entanglement entropy is expected to be extensive, and
we introduce the entropy density
\begin{equation}
  s= \frac{S}{L}.
\end{equation}
It is expected that $s$ should not depend on $L$ apart from minor finite size effects.

\begin{figure*}[!ht]
\minipage{0.33\textwidth}
\includegraphics[width=\linewidth]{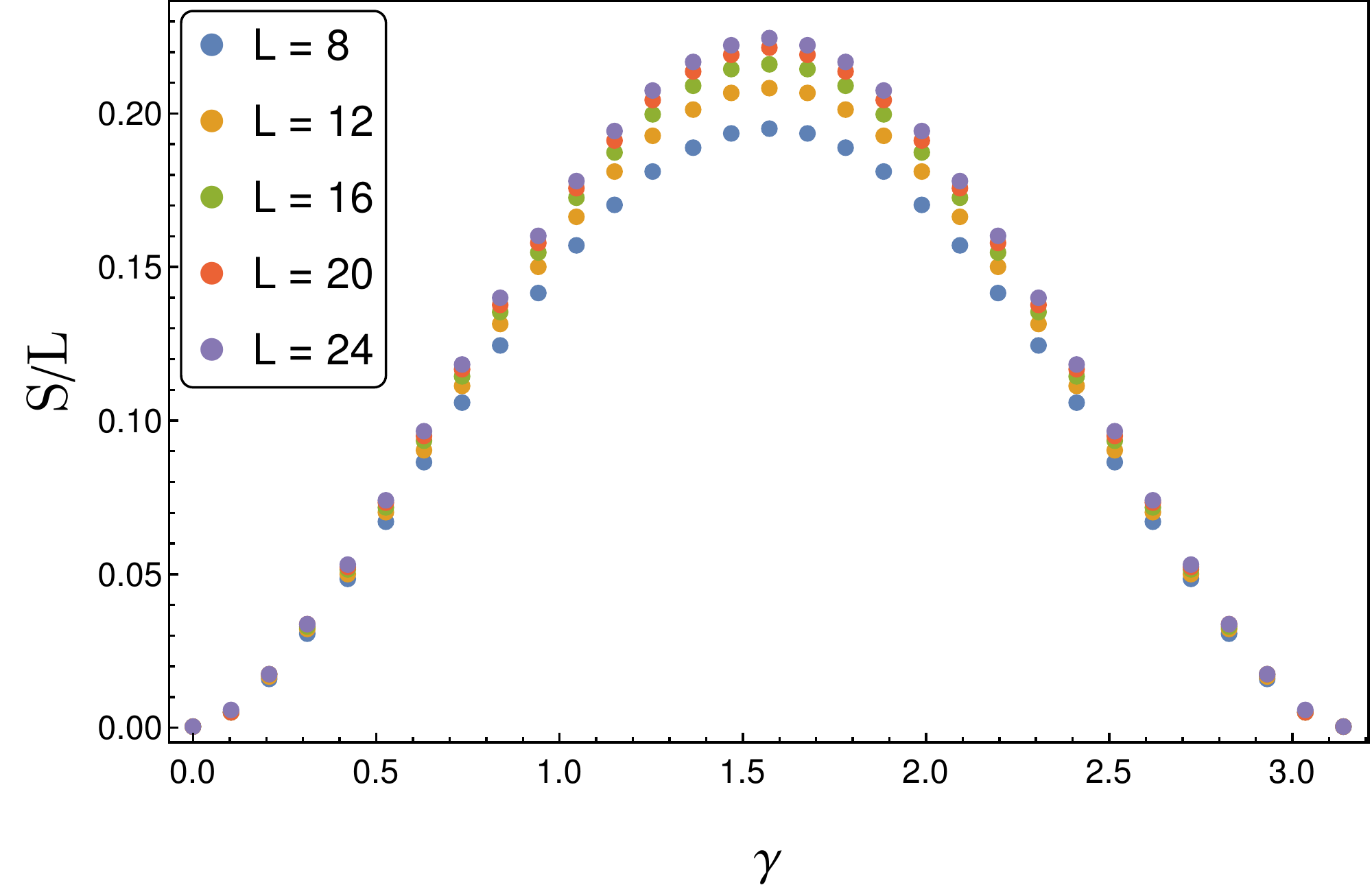}  
\endminipage\hfill
\minipage{0.33\textwidth}
\includegraphics[width=\linewidth]{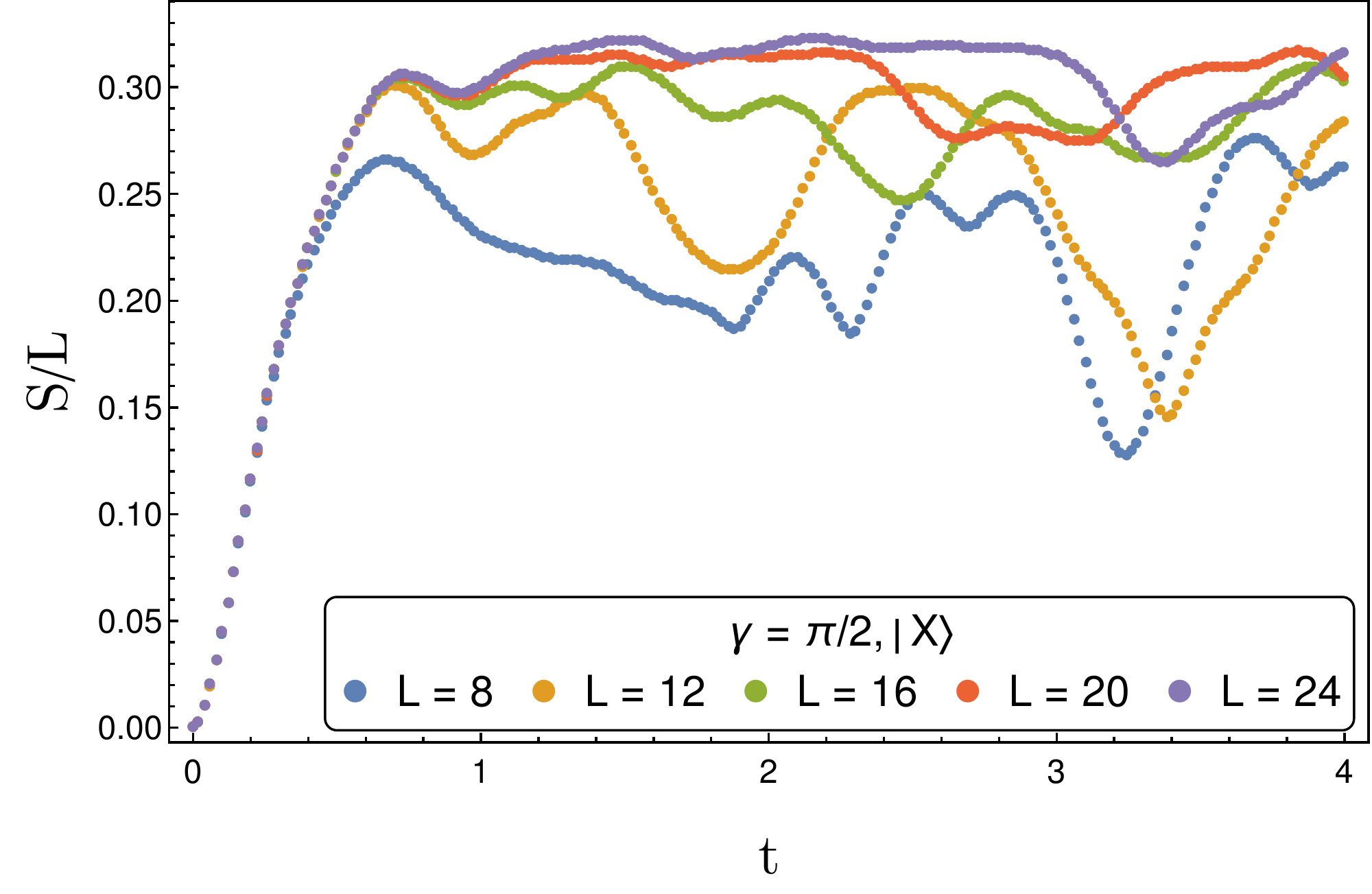}  
\endminipage\hfill
\minipage{0.33\textwidth}
\includegraphics[width=\linewidth]{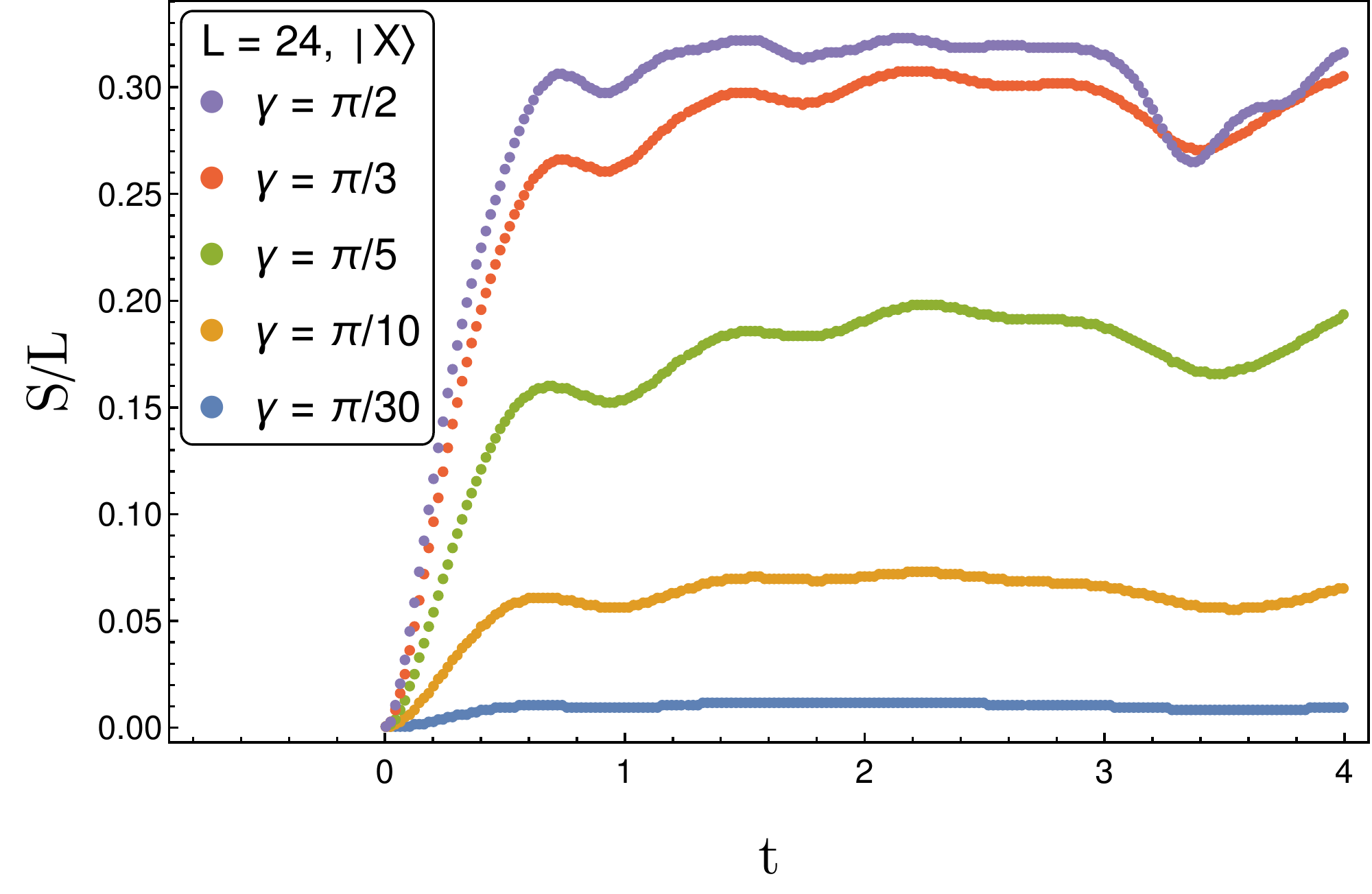}  
\endminipage
\caption{Numerical results for the sub-lattice entanglement density $s=S/L$. Left: Ground state values as a function of
  the coupling $\gamma$ for different values of $L$. Middle: Time evolution of the entanglement in a quench problem (see
  main text), for
  different volumes and $\gamma=\pi/2$. Right: Time evolution in a quench problem for $L=24$ and different values of $\gamma$.}
\label{fig:2}
\end{figure*}

We numerically study $s$ in the ground state for different values of $\gamma$ and $L$, results are plotted
in Fig. \ref{fig:2}. We find that finite size effects are indeed small,  and the entropy density obtains its maximum
value at the free fermion point $\gamma=\pi/2$. 

We also consider non-equilibrium time evolution started from a selected initial state, namely the
ferromagnetic state with polarization in the $x$-direction:
\begin{equation}
   \ket{X}=\bigotimes_{j=1}^L  \frac{1}{\sqrt{2}} \Big(\ket{\uparrow}+\ket{\downarrow}\Big).
\end{equation}
We study the real time evolution of $s$, results are plotted in Fig \ref{fig:2}. 
It can be seen that the entanglement is indeed extensive, but now there are bigger finite size effects.
For the largest system size $L=24$ we see that entanglement reaches a plateau relatively soon for all values of $\gamma$, with
the height of the plateau behaving in a similar way as before: the maximum entanglement is seen for
$\gamma=\pi/2$. Qualitatively similar behaviour can also be found for other initial states. 

\section{Discussion}

We introduced a new exactly solvable spin ladder,
which is one of the simplest
quantum integrable models with a tunable coupling between particles.
The model interpolates between two free fermion points, in which the two sub-lattices are either uncoupled or
maximally coupled. 
The entanglement between the sub-lattices was demonstrated analytically by an exact result for a correlation function,
and numerically by  the sub-lattice entanglement
\eqref{Sdef} defined in the real space basis, which was examined both in and out of equilibrium. 

In the special case $\gamma=\pi/2$ the entanglement can
also be studied in terms of the fermionic degrees of freedom. It was pointed out in \cite{igloi-peschel}
that for free fermionic chains these two definitions of entanglement give generally different results. They only agree 
for connected sub-systems, because for disconnected sub-systems the
Jordan-Wigner transformation causes differences between the two types of entanglement. Our model can be regarded as an
extreme example for this phenomenon: for $\gamma=\pi/2$ the two sub-lattices are completely decoupled if one considers
the fermions, see eq. \eqref{Hc}. Therefore, the sub-lattice entanglement in terms of the fermions is exactly zero. In
contrast, we find that the real space entanglement is non-zero, and in fact it is maximal for the free fermion point $\gamma=\pi/2$! 

Finally we note that the model is partially anyonic for a generic $\gamma$, and it seems to be one of the simplest non-trivial scattering theories.
This could lead to interesting applications, for example in the realization of interacting Bethe states in
quantum computers \cite{bethe-qcomp-1,bethe-qcomp-2,bethe-qcomp-3,bethe-qcomp-4,bethe-qcomp-5}.

\vspace{0.5cm}

{\it Note added:} After this work was completed we became aware of the recent work \cite{anyon-ladder-1} which treats a closely
related model. Our results about  the integrability and exact solvability of the model (together with the exact result for a
correlation function) appear to be new, whereas the results for sub-lattice entanglement are partly overlapping.

\vspace{0.5cm}

\begin{acknowledgments}
  We acknowledge inspiring discussions with Lorenzo Piroli, Eric Vernier and Viktor Eisler, and we are thankful to
Bal\'azs D\'ora,  Paul  Fendley, Sarang Gopalakrishnan, Yuan Miao and Romain Vasseur for discussions about existing
literature, and we thank Adhip Agarwala for bringing the work \cite{anyon-ladder-1} to our attention.
The work of GT has been supported by the National Research Development and Innovation Office (NKFIH) through Grant Nos. SNN139581 and K138606, and within the Quantum Information National Laboratory. LP acknowledges support from the ÚNKP-21-3-II New National Excellence Program of the Ministry of Innovation and Technology.
\end{acknowledgments}

\end{document}